\documentclass[useAMS,usenatbib,usegraphicx]{mn2e}
\usepackage{graphicx}
\usepackage{color}
\usepackage{amssymb}

\title[Determining physical parameters of RRab stars]
{A new method for determining physical parameters of fundamental mode RR~Lyrae stars from multicolour light curves}
\author[\'A. S\'odor et al.]{
\'A. S\'odor, J. Jurcsik, and B. Szeidl
\\
Konkoly Observatory of the Hungarian Academy of Sciences, H--1525 Budapest PO Box 67, Hungary\\
}
\begin{document}

\date{Accepted 2008 November 18.  Received 2008 November 10; in original form 2008 July 24}

\pagerange{\pageref{firstpage}--\pageref{lastpage}} \pubyear{2008}

\maketitle

\label{firstpage}

\begin{abstract}
We present a new method for determining physical parameters of RRab variables exclusively from multicolour light curves. Our method is an Inverse Photometric Baade-Wesselink analysis which, using a nonlinear least squares algorithm, searches for the effective temperature ($T_\mathrm{eff}$) and pulsational velocity ($V_\mathrm{p}$) curves and other physical parameters that best fit the observed light curves, utilising synthetic colours and bolometric corrections from static atmosphere models. The $T_\mathrm{eff}$ and $V_\mathrm{p}$ curves are initially derived from empirical relations then they are varied by the fitting algorithm. The method yields the variations and the absolute values of the radius, the effective temperature, the visual brightness, and the luminosity of individual objects. Distance and mass are also determined. The method is tested on 9 RRab stars subjected to Baade-Wesselink analyses earlier by several authors. The physical parameters derived by our method using only the light curve data of these stars are well within their possible ranges defined by direct Baade-Wesselink and other techniques. A new empirical relation between the $I_\mathrm{C}$ magnitude and the pulsational velocity is also presented, which allows to construct the $V_\mathrm{p}$ curve of an RRab star purely from photometric observations to an accuracy of about 3.5\,km/s.
\end{abstract}


\section{Introduction}

The light variations of pulsating variable stars contain much useful information both on the physics of the brightness changes and also on the global parameters of the objects. Concerning RR~Lyrae stars, this information can be retrieved using e.g., different variants of the Baade-Wesselink (\hbox{B-W}) analysis \citep{baade,wesselink,mandbell,gautschy,fernley90} or empirical relations between the Fourier parameters of the light curves and certain physical parameters of the variables \citep[e.g.][]{jk96,jurcsik98,kw}. However, there are cases, for example, Blazhko RR~Lyrae stars showing light curve variation when we are encountering problems with both methods. The empirical light curve -- physical parameter relations are proved to be valid only for RR~Lyrae stars with reasonably stable light curves i.e., for unmodulated or weakly modulated stars. A good example for the latter is RR~Gem, which is one of the calibrator objects of these relations, while its light curve is weakly but unambiguously modulated \citep{rrgem1, rrgem2}. In spite of its small amplitude light-curve modulation, RR Gem fits well all the empirical light curve -- physical parameter relations. The Fourier parameters of the mean pulsation light curve of a large modulation amplitude Blazhko star, however, differ significantly from the parameters calculated for the light curves at any phase of the modulation (see e.g. MW~Lyr, \citealt{mwlyr}), which makes the inference of the empirical relations ambiguous for these Blazhko variables. Applying the empirical formulae on the Fourier parameters of MW Lyr in different phases of the modulation \cite[data from Table 7 in][]{mwlyr} variations of 0.45 dex in [Fe/H] are obtained during the Blazhko cycle. It would be difficult to find a physical explanation for such a metallicity variation. Moreover, the magnitude averaged absolute brightness changes calculated from the Fourier parameters are 0.025 mag larger than what is in fact directly observed. Therefore, at present the light curve-physical parameter formulae are not appropriate to follow the changes in the mean physical parameters during the Blazhko cycle.

The \hbox{B-W} analysis is capable for determining physical parameters of Blazhko stars for any phase of the modulation if simultaneous $V_\mathrm{rad}$ and multicolour photometric observations were available. Thus, any changes in the mean physical parameters could be, in principle, detected. However, a practical problem is that for obtaining such results simultaneous extended spectroscopic radial velocity and multicolour photometric observations were needed with good phase coverage of both the pulsation and the modulation cycles. For such an investigation at least one order of magnitude more $V_\mathrm{rad}$ data points were needed than are available today.

Up to now, there are only about two dozens of unmodulated field RRab stars, about one dozen RRab stars in globular clusters, and no Blazhko variable which have simultaneous photometric and $V_\mathrm{rad}$ observations suitable for \hbox{B-W} analysis. On the contrary, extended and accurate multicolour observations of many modulated and unmodulated RRab stars have become available from our CCD observations at the Konkoly Observatory in the past several years \citep{sodorAN}. To extract physical information exclusively from multicolour light curves of RRab stars we have developed a method that fits the observed magnitudes and colours with appropriate physical parameters and their variations. The main advantage of the method is that all these are derived exclusively from photometric observations without any need of spectroscopic measurements.

In this paper our method is introduced and is tested on 9 non-modulated RRab stars. The application of the method on a large modulation amplitude RR~Lyrae star, MW~Lyr \citep{mwlyr}, is presented in \cite{mwlyr2}.

\section{The method}

The basic idea of the Inverse Photometric Baade-Wesselink method (IP method) is to look for the pulsational velocity ($V_\mathrm{p}(\varphi)$) and effective temperature ($T_\mathrm{eff}(\varphi)$) curves ($\varphi$ denotes the pulsation phase), the distance, and the mean radius of the star which best reproduce the input light curves utilising static atmosphere model results.

The inputs of the method are the Fourier series fitted to the multicolour $BVI_\mathrm{C}$ light curves and the metallicity ([Fe/H]). The input Fourier curves may represent either the light curves of an unmodulated RRab star, the mean pulsation light curves of a modulated star, or the light curves of a modulated RR Lyrae star at a certain Blazhko phase. To obtain correct results the extinction corrected apparent visual brightness and the intrinsic colours have to be known. However, as it is shown in Sect.~\ref{sect:app-zp} the latter requirement is not always crucial, since the colour zero points can be determined by the method itself if good quality light curves are available.

For using model atmosphere tables the value of [Fe/H] and the instantaneous values of $T_\mathrm{eff}(\varphi)$ and the effective gravity ($\log g_\mathrm{eff}(\varphi)$) are needed. The latter is determined from the $V_\mathrm{p}(\varphi)$ curve, the mass ($\mathfrak{M}$), and the average radius ($R_0$) of the star, as follows. The instantaneous radius ($R(\varphi)$) and the radial acceleration ($a(\varphi)$) are computed from $V_\mathrm{p}(\varphi)$ as the first integral and the first derivative, respectively. Knowing these variables, $g_\mathrm{eff}(\varphi)$ are then calculated according to the formula:

\begin{equation}
g_\mathrm{eff}(\varphi) = G\mathfrak{M}/R(\varphi)^2 + a(\varphi)
\end{equation}

\noindent where $G$ is the gravitational constant. The mass satisfies the modified form of the pulsation equation:
 
\begin{equation}
\log \mathfrak{M} = -2.237 + 2.56 (\log R)_0 - 1.538 \log P - 0.054 (\log T_\mathrm{eff})_0 \label{eq:puls}
\end{equation}

\noindent which has been derived from Eq.~1a of \cite{marconi03} by substituting the mean logarithm of the luminosity, $(\log L)_0$ with:

\begin{equation}
(\log L)_0 = 4 (\log T_\mathrm{eff})_0 + 2 (\log R)_0 - 15.048.
\end{equation}

\noindent Note that $\mathfrak{M}$, $L$, and $R$ are in Solar units and `0' index denotes pulsation-cycle-averaged values. $P$ is the pulsation period of the variable in days.

Synthetic colours, $(B-V)(\varphi)$ and $(V-I_\mathrm{C})(\varphi)$, and bolometric correction, $BC(\varphi)$, corresponding to [Fe/H],  $T_\mathrm{eff}(\varphi)$ and $\log g_\mathrm{eff}(\varphi)$ are taken from the static atmosphere model tables of \cite{cast}. We applied linear interpolation in [Fe/H] to obtain a resolution of 0.05\,dex between the 0.5\,dex steps of the original tables. The original resolution of the $T_\mathrm{eff}$ and $\log g$ grid are 250\,K and 0.5\,dex, respectively, which have been interpolated by spline functions. The extinction corrected apparent magnitudes of the star is then derived from:
\begin{equation}
V(\varphi) = -10 \log T(\varphi) - 5 \log R(\varphi) - BC(\varphi) + 5 \log d + 37.36
\end{equation}

\noindent where $d$ is the distance in pc.

An implementation of the Levenberg-Marquardt nonlinear fitting algorithm by \cite{lourakis04LM} is applied to fit the derived light and colour curves to the observations with the alterations of the initial $T_\mathrm{eff}(\varphi)$ and $V_\mathrm{p}(\varphi)$ curves, $d$, and $R_0$. To obtain physically plausible results continuity, smoothness, and periodicity of the variations of the physical parameters have to be maintained during the fitting process. In order to attain these, the initial  $T_\mathrm{eff}(\varphi)$ and $V_\mathrm{p}(\varphi)$ functions are defined as  smooth and continuous funcions which are then modified by the fitting algorithm with the addition of appropriate order Fourier series.

RR Lyr atmospheres differ from static atmospheres in many respects. Just to mention some of the most problematic issues, in the dynamic atmosphere of large amplitude pulsating variables shock waves propagate, local thermodynamic equilibrium does not fulfil (NLTE), the atmosphere is not thin compared to the radius of the star while plan-parallel models are calculated, the turbulent velocity ($v_\mathrm t$) also varies during the pulsation cycle, etc. However, as atmosphere models taking into account all these complications (or at least some of them) have not been calculated yet for a wide enough range of parameters, all the methods which are applied today to define the physical parameters of pulsating variables from photometric and spectroscopic data relies on static atmosphere models to some extent. The applicability of static atmosphere models for dynamic atmospheres of pulsating RR~Lyrae stars has already been discussed by many authors \citep[e.g.][]{lj90, jones92}. It has been shown that the vicinity of the minimum radius phase is that part of the light curve where the validity of static models is the most questionable. Notwithstanding, in the IP method the whole phase range of the pulsation is used in order to maintain the continuity of the curves. The smoothness and continuity of the curves are especially important because, due to the dynamic nature of the atmosphere, and to possible differences of the temperature scales the different colours corresponding the input light curves are sometimes out of the ranges of the static atmosphere grid e.g., on the \hbox{$(B-V)$--$(V-I_\mathrm{C})$} plane which would otherwise introduce discontinuity in the output curves. Since we do not use the input colour and magnitude values at any phase directly for calculating physical quantities, problems that arise from inappropriateness of the static atmosphere models at certain phases does not affect our results directly. Our inverse method ensures that the solution curves are completely within the ranges of the model grid. Where the static model is less appropriate a poorer fit is expected.

\subsection{The initial $T_\mathrm{eff}(\varphi)$ curve}

The $(V-I_\mathrm{C})$ colour is a good temperature indicator both theoretically and empirically. We use the polynomial transformation given by \citet[Table 8.]{bessel} to calculate an initial $T_\mathrm{eff}(\varphi)$ function from the input $(V-I_\mathrm{C})(\varphi)$ curve. The actual accuracy of this formula is not an issue because a number of test runs show that the fitting process converges to the same solution $T_\mathrm{eff}(\varphi)$ curve when started from a wide range of initial functions i.e., the results are not sensitive to its choice. The solution $T_\mathrm{eff}(\varphi)$ curve depends slightly on the choice of the initial $V_\mathrm{p}(\varphi)$ function, but its effect is less than $10-30$\,K in any phase of the pulsation.

\subsection{The initial $V_\mathrm{p}(\varphi)$ curve}

In contrast with the stability of the $T_\mathrm{eff}(\varphi)$ function, the situation with the pulsational velocity, $V_\mathrm{p}(\varphi)$, is quite different. We have found that, during the fitting process, the $V_\mathrm{p}(\varphi)$ curve departs from a physically plausible shape if the fitting method is allowed to change its shape arbitrarily. Therefore, it is essential to find a reasonably good initial $V_\mathrm{p}(\varphi)$ curve and, to get a reliable result, the solution $V_\mathrm{p}(\varphi)$ curve has to be kept near to the initial one.

\subsubsection{The $V_\mathrm{p}(\varphi)$ curve given by Liu (1991)}
\label{sect:liuvrad}

\cite{liu91} published a template $V_\mathrm{p}(\varphi)$ curve of RRab stars based on about a dozen observed $V_\mathrm{rad}(\varphi)$ curves. The amplitudes of the input curves were normalized to unity and they were phased to coincide at their minima. The average of these curves was then calculated to define the shape of a template $V_\mathrm{p}(\varphi)$ curve. \cite{liu91} also derived a relation which gave the amplitude of the $V_\mathrm{p}(\varphi)$ curve as a function of the $V(\varphi)$ light curve amplitude with an accuracy of about 5\,km/s. The actual template $V_\mathrm{p}(\varphi)$ curve of an RRab variable is defined by the template shape and this amplitude formula. This template $V_\mathrm{p}(\varphi)$ curve describes the input curves with about 5\,km/s accuracy, except on the steep descending branch between phases 0.85 and 1.0, where the accuracy is only about 25\,km/s. 

This average curve has, however, several drawbacks. Due to phase differences on the very steep descending branch, the template curve is less steep here as any of the individual curves \cite[see Fig.~2 in][]{liu91}. For similar reason, the minimum is less sharp than most of the individual $V_\mathrm{p}(\varphi)$ curves.

Another problem with the template $V_\mathrm{p}(\varphi)$ curve of \cite{liu91} is its fixed shape. Our main purpose in developing the IP method was to study Blazhko stars showing large amplitude light curve modulation. Looking at the large variety of the light curves' shape of a Blazhko variable during the Blazhko cycle, it definitely does not seem to be a good decision to fix the shape of the $V_\mathrm{p}(\varphi)$ curve to a unique template. Therefore, we have been looking for some connection between the $V_\mathrm{p}(\varphi)$ curve, and the colour and light curves.

\subsubsection{$V_\mathrm{p}(\varphi)$ curve calculated from the $I_\mathrm{C}(\varphi)$ light curve}
\label{sect:i-vrad}

The only $V_\mathrm{rad}$ observation that covers both the pulsation and modulation cycles of a Blazhko star was obtained by \cite{chadid}. Unfortunately, there is no simultaneous spectroscopic and photometric observation of any Blazhko star that would permit to examine the possibility of a light curve -- $V_\mathrm{p}$ curve connection valid for Blazhko stars in the course of the modulation. Therefore, in the following we consider only unmodulated RRab stars.

We used the photometric and $V_\mathrm{rad}$ data of RRab stars collected by \cite{kovacs}. From Kov\'acs's sample of 22 RR~Lyrae stars 20 are utilised, as RR~Leo and W~Crt has been omitted because of their discrepant curves. Additionally, we also used data of RV~Oct. This sample contains all the RR~Lyrae stars that have accurate $V_\mathrm{rad}$ and photometric observations with good phase coverage of the pulsation.

Following \cite{kovacs} we use $p=1.35$ projection factor to transform radial velocities to pulsational velocities according to the formula:

\begin{equation}
V_\mathrm{p} = p \cdot V_\mathrm{rad},\ \ \mathrm{where}\ \ \ p=1.35.
\end{equation}

\begin{figure} 
\begin{center}
\includegraphics[angle=0,width=8cm]{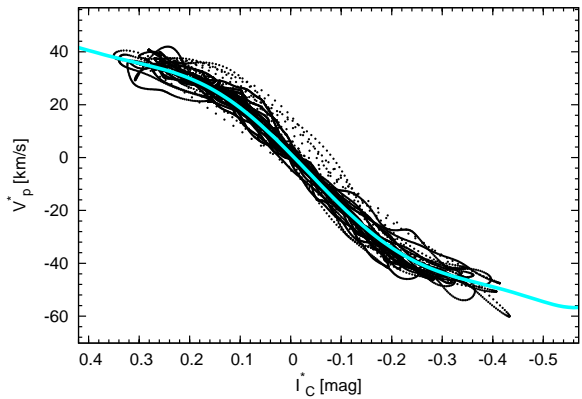}
\end{center}
\caption{
$V_\mathrm{p}^{*}(\varphi)$ vs.\ $I_\mathrm{C}^{*}(\varphi)$ relation. The plotted values are calculated from the Fourier fits of the light and radial velocity curves of 21 stars sampled with 1/500 phase resolution. Asterisks denote that the time averages are subtracted from the data. The solid grey (cyan) line shows a 7th order polynomial fit according to Eq.~\ref{eq:i-vrad}.\label{fig:i-vrad}
}
\end{figure}

Inspecting the observed $V_\mathrm{rad}(\varphi)$ and light and colour curves, we have found that the shape of the $I_\mathrm{C}(\varphi)$ curve resembles the most to that of the $V_\mathrm{rad}(\varphi)$ curve. The $V_\mathrm{p}^{*}$ vs.\  $I_\mathrm{C}^{*}$ values of the 21 RRab stars are plotted in Fig.~\ref{fig:i-vrad}. Asterisk denotes that their phase-averaged value is subtracted from the data i.e., they are shifted to the same  $V_\mathrm{p}$ and magnitude zero points. This plot shows that the $V_\mathrm{p}^{*}(\varphi)$ of an RRab star can be expressed as a function of its $I_\mathrm{C}^{*}(\varphi)$ brightness at any phase of the pulsation with sufficient accuracy.

\begin{figure} 
\begin{center}
\includegraphics[angle=0,width=6cm]{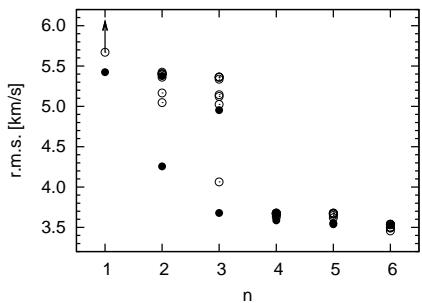}
\end{center}
\caption{$r.m.s.$ values of the regressions of $V_\mathrm{p}^{*}$ with $I_\mathrm{C}^{*k}$, $(V-I_\mathrm{C})^l$, $(B-V)^l$,  $I_\mathrm{C}^{*k}\cdot(V-I_\mathrm{C})^l$, and $I_\mathrm{C}^{*k}\cdot(B-V)^l$ terms. Filled circles denote the $n$th degree regressions when only $I_\mathrm{C}^{*k}$ terms are used. Open circles denote regressions where $n-1$ terms of $I_\mathrm{C}^{*k}$ and one colour or compound term are taken into account. The $r.m.s.$ does not drop significantly after the 3rd degree. The best 3rd order solution is achieved by a regression using the $I_\mathrm{C}^{*}$, $I_\mathrm{C}^{*2}$, and $I_\mathrm{C}^{*3}$ terms. The arrow at the top left corner shows that data points here are at high $r.m.s.$ values, between 9\,km/s and 20\,km/s.\label{fig:i-vrad-reg}}
\end{figure}

Statistical tests were performed to determine on how many parameters $V_\mathrm{p}^{*}$ depends significantly and whether it is sufficient to express it with $I_\mathrm{C}^{*k}$ terms or it also depends on the colours. The $r.m.s.$ of different degree regressions of $V_\mathrm{p}^{*}$ with terms of $I_\mathrm{C}^{*k}$, $(V-I_\mathrm{C})^l$, $(B-V)^l$, $I_\mathrm{C}^{*k}\cdot(V-I_\mathrm{C})^l$, and $I_\mathrm{C}^{*k}\cdot(B-V)^l$ are plotted in Fig.~\ref{fig:i-vrad-reg}. The plot reveals that no significant improvement can be achieved when we take into account a colour or compound term, too. It can also be seen in Fig.~\ref{fig:i-vrad-reg} that a 3rd order polynomial fit with $I_\mathrm{C}^{*}$, $I_\mathrm{C}^{*2}$, and $I_\mathrm{C}^{*3}$ terms describes $V_\mathrm{p}^{*}$ with sufficient accuracy. However, if we fit the data directly by this polynomial, it will not be monotonic at the ends of the $I_\mathrm{C}^{*}$ range covered by the data. Therefore, to obtain a strictly monotonic \hbox{$V_\mathrm{p}^{*}$ -- $I_\mathrm{C}^{*}$} relation, the following steps were performed. The plotted data were smoothed with a cubic spline and the spline was extrapolated linearly over the boundaries of the covered $I_\mathrm{C}^{*}$ range with about 0.05\,mag at both ends. This spline was then fitted with a 7th degree polynomial.
Finally, we found the following relation between $V_\mathrm{p}^{*}$ and $I_\mathrm{C}^{*}$, which represents the $V_\mathrm{p}^{*}(\varphi)$ curve of RRab stars similarly good or even better than Liu's template does. The formula:

\begin{equation}
V_\mathrm{p}^{*}(\varphi) = \sum_{i=0}^{7} c_\mathrm{i} \cdot I_\mathrm{C}^{*}(\varphi)^\mathrm{i} \label{eq:i-vrad}
\end{equation}
\noindent where $c_\mathrm{0..7} = (1.38, 194.8, -114.5, -1023,$

\hskip15mm $ 764, 4220, -1660, -6700)$

\noindent fits the observations of the 21 stars with 3.51\,km/s accuracy. The errors of the coefficients are:

\noindent \hskip10mm $\sigma(c_\mathrm{0..7}) = (0.08, 0.7, 3.7, 14, 40, 100, 110, 200)$

\vskip24pt

When running the IP code, initial $V_\mathrm{p}(\varphi)$ curves derived either from Liu's template or from the \hbox{$V_\mathrm{p}^{*}$ -- $I_\mathrm{C}^{*}$} relation are applied. As we are not sure that any of these $V_\mathrm{p}(\varphi)$ curves are correct indeed, the shapes of the solution $V_\mathrm{p}(\varphi)$ curves are not fixed to these templates. Unfortunately, as we mentioned earlier the $V_\mathrm{p}(\varphi)$ curve converges to a physically irrealistic solution if it is allowed to change arbitrarily. Therefore the fitting algorithm treats the initial $V_\mathrm{p}(\varphi)$ template as another curve that is fitted with an actual, varied $V_\mathrm{p}(\varphi)$ curve. In this way, it becomes possible to give penalties to the trials depending on how much their $V_\mathrm{p}(\varphi)$ curve deviates from the initial one. The rate of this penalty can be altered by a weight factor. Two weight factors are used. One allows $V_\mathrm{p}(\varphi)$ to depart from the initial curve with about $4$\,km/s or less in any phase of the pulsation (medium weight). The other weight is 100 times larger, allowing practically no change in $V_\mathrm{p}(\varphi)$ (large weight). The variation of the initial $V_\mathrm{p}(\varphi)$ curve is allowed with the addition of a 2nd order Fourier series, as our experiments show that higher orders lead to wavy solution curves. However, the results of the IP method depend more significantly on the quality of the input light curves than on the choice and variance of the initial $V_\mathrm{p}(\varphi)$ curves. The $3-5$\,km/s accuracy of the initial $V_\mathrm{p}(\varphi)$ curves makes it possible to achieve similarly accurate results with the IP method using exculsively good quality photometric data as with earlier direct \hbox{B-W} analyses.

As our method fits the output mathematical curves to input mathematical curves, the formal errors of the derived parameters according to the Levenberg-Marquardt fitting algorithm are meaningless and are independent from the uncertainties of the underlying observations. However, the results of the IP method are very sensitive to the accuracy and phase coverage of the photometric observations, since for good quality data the results depend only slightly on the internal settings of the code ($V_\mathrm{rad}$ template, weight factor) while for worse data quality the results can show relatively large dispersion. Another problem of estimating the errors of the parameters derived from the IP method is that accurate fits of the input data can also be obtained for physically unreliable conditions, e.g, when the mass is not fixed by the pulsation equation, or when fitting the initial radial velocity curve has only a small weight. Therefore we regard the range of the output values from four different runs with different settings listed in Table 1 as an estimate of the inherent uncertainty of the IP method.

\begin{table} 
\caption{Applied settings of the IP method.\label{tbl:abcd}}
\begin{center}
\begin{tabular}{ccc}
\hline
notation & initial $V_\mathrm{p}(\varphi)$ template & $V_\mathrm{p}(\varphi)$ weight factor \\
\hline
A & $V_\mathrm{p}^{*}$ -- $I_\mathrm{C}^{*}$ (Eq.~\ref{eq:i-vrad}) & medium \\
B & $V_\mathrm{p}^{*}$ -- $I_\mathrm{C}^{*}$ (Eq.~\ref{eq:i-vrad}) & large \\
C & \cite{liu91} & medium \\
D & \cite{liu91} & large \\
\hline
\end{tabular}
\end{center}
\end{table}

\section{Test results}

\begin{table*} 
\caption{Comparison of the results of the IP method with literature data for the 9 test objects. The settings of the IP method using the notation of Table~\ref{tbl:abcd} are given in the `remark/ref.' column. Italic numbers denote  photometric metallicities and values that were calculated in this paper from the other parameters according to Eqs.~\ref{eq:puls}, \ref{eq:logg}, and \ref{eq:logd}.\label{tbl:res}}
\begin{tabular}{lp{15mm}cccccc}
\hline
GCVS name $(\log P)$& [Fe/H] & $M_\mathrm{V\,0}$ & $d$ & $R_0$       & $T_\mathrm{eff\,0}$ & $\mathfrak{M}$         & $\log g_\mathrm{stat}$ \\
\hskip5mm remark/ref.        &        & [mag]             & [pc]& [$R_\odot$] & [K]                 & [$\mathfrak{M}_\odot$] \\
\hline\hline
\multicolumn{4}{l}{SW And (-0.354)}\\
\hline
\hskip5mm A                             & \it{-0.10} & 1.01 &\ 504 & 4.34 & 6646 & 0.56 & 2.90 \\
\hskip5mm B                             & \it{-0.10} & 1.00 &\ 507 & 4.41 & 6645 & 0.57 & 2.90 \\
\hskip5mm C                             & \it{-0.10} & 0.99 &\ 508 & 4.42 & 6646 & 0.57 & 2.90 \\
\hskip5mm D                             & \it{-0.10} & 0.97 &\ 513 & 4.46 & 6644 & 0.58 & 2.90 \\
\hline
\hskip5mm \cite{mcnamara}               &\ 0.00 & 0.90 &\      \it{532} & 4.45 & 6680 & 0.60 & 2.92 \\
\hskip5mm \cite{cacc89}                 & -0.15 & 0.88 &\      \it{537} & 4.49 & 6640 & 0.60 & 2.91 \\
\hskip5mm \cite{lj90}                   & -0.10 & 0.97 &\          511 & 4.36 & 6517 & 0.56 & \it{2.91} \\
\hskip5mm \cite{jones92}                & -0.15 & 1.10 &\          520 & 4.06 \\
\hskip5mm \cite{fernley}                & -0.15 & 0.94 &\      \it{522} & \\
\hskip5mm \cite{kovacs}                 & \it{-0.10} & 0.81 &\ \it{555} & 4.47 & 6702 & \it{0.58} & \it{2.90} \\
\hskip5mm \cite{bono}                   & -0.24 & 0.93 &\      \it{523} \\
\hline\hline
\multicolumn{4}{l}{WY Ant (-0.241)}\\
\hline
\hskip5mm A                             & \it{-1.55} & 0.36 & 1190 & 6.38 & 6527 & 0.97 & 2.82 \\
\hskip5mm B                             & \it{-1.55} & 0.34 & 1200 & 6.44 & 6526 & 0.99 & 2.82 \\
\hskip5mm C                             & \it{-1.55} & 0.52 & 1105 & 5.93 & 6525 & 0.81 & 2.80 \\
\hskip5mm D                             & \it{-1.55} & 0.49 & 1119 & 6.00 & 6525 & 0.83 & 2.80 \\
\hline
\hskip5mm \cite{skillen}                & -1.25 & 0.63 &          1035 & 5.61 & 6389 & 0.72 & \it{2.80} \\
\hskip5mm \cite{fernley}                & -1.25 & 0.55 &      \it{1089} &      &      &      &      \\
\hskip5mm \cite{kovacs}                 & \it{-1.55} & 0.52 & \it{1105} & 6.10 & 6296 & \it{0.87} & \it{2.81} \\
\hskip5mm \cite{bono}                   & -1.48 & 0.48 &      \it{1124} \\
\hline\hline
\multicolumn{4}{l}{UU Cet (-0.217)}\\
\hline
\hskip5mm A                             & \it{-1.10} & 0.66 & 1865 & 5.72 & 6420 & 0.68 & 2.75 \\
\hskip5mm B                             & \it{-1.10} & 0.62 & 1896 & 5.82 & 6419 & 0.71 & 2.76 \\
\hskip5mm C                             & \it{-1.10} & 0.49 & 2016 & 6.19 & 6420 & 0.83 & 2.78 \\
\hskip5mm D                             & \it{-1.10} & 0.44 & 2061 & 6.32 & 6420 & 0.88 & 2.78 \\
\hline
\hskip5mm \cite{cacc92} IR Flux method  & -1.00 & 0.63 &          1887 & 5.75 & 6300 & 0.70 & \it{2.76} \\
\hskip5mm \cite{cacc92} SB method       & -1.00 & 0.70 &          1825 & 5.36 &      & 0.59 & \it{2.75} \\
\hskip5mm \cite{cacc92} SB method       & -1.00 & 0.52 &          1982 & 5.50 &      & 0.62 & \it{2.75} \\
\hskip5mm \cite{fernley}                & -1.00 & 0.62 &      \it{1899} &      &      &      &      \\
\hskip5mm \cite{kovacs}                 & \it{-1.10} & 0.48 & \it{2025} & 6.25 & 6258 & \it{0.85} & \it{2.77} \\
\hskip5mm \cite{bono}                   & -1.28 & 0.56 &      \it{1949} \\
\hline\hline
\multicolumn{4}{l}{SU Dra (-0.180)}\\
\hline
\hskip5mm A                             & \it{-1.60} & 0.60 & \ 695 & 5.91 & 6431 & 0.63 & 2.70 \\
\hskip5mm B                             & \it{-1.60} & 0.55 & \ 712 & 6.05 & 6428 & 0.69 & 2.71 \\
\hskip5mm C                             & \it{-1.60} & 0.60 & \ 696 & 5.91 & 6432 & 0.65 & 2.70 \\
\hskip5mm D                             & \it{-1.60} & 0.51 & \ 724 & 6.15 & 6431 & 0.71 & 2.71 \\
\multicolumn{2}{l}{\hskip5mm A$_\alpha$\hskip25mm $[{\rm M}/{\rm H}]=-1.6$; $[\alpha/{\rm Fe}]=0.4$}
                                                     & 0.57 & \ 703 & 5.96 & 6435 & 0.66 & 2.71 \\
\multicolumn{2}{l}{\hskip5mm B$_\alpha$\hskip25mm $[{\rm M}/{\rm H}]=-1.6$; $[\alpha/{\rm Fe}]=0.4$}
                                                     & 0.51 & \ 723 & 6.13 & 6433 & 0.71 & 2.71 \\
\multicolumn{2}{l}{\hskip5mm C$_\alpha$\hskip25mm $[{\rm M}/{\rm H}]=-1.6$; $[\alpha/{\rm Fe}]=0.4$}
                                                     & 0.56 & \ 707 & 5.99 & 6437 & 0.67 & 2.71 \\
\multicolumn{2}{l}{\hskip5mm D$_\alpha$\hskip25mm $[{\rm M}/{\rm H}]=-1.6$; $[\alpha/{\rm Fe}]=0.4$}
                                                     & 0.47 & \ 737 & 6.24 & 6438 & 0.74 & 2.72 \\
\hline
\hskip5mm \cite{lj90}                   & -1.60 & 0.73 &          \ 640 & 5.15 & 6433 & 0.47 & \it{2.69} \\
\hskip5mm \cite{fernley}                & -1.60 & 0.63 &      \ \it{685} &      &      &      &      \\
\hskip5mm \cite{barcza}                 & -1.60 & 0.74 &          \ 647 & 5.09 & 6490 & 0.66 & \it{2.84} \\
\hskip5mm \cite{kovacs}                 & \it{-1.60} & 0.62 & \ \it{689} & 5.85 & 6293 & \it{0.63} & \it{2.70} \\
\hskip5mm \cite{bono}                   & -1.80 & 0.26 &\       \it{813}\\
\hline\hline
\end{tabular}
\end{table*}

\begin{table*} 
\caption{Results (continued).\label{tbl:res2}}
\begin{tabular}{lp{15mm}cccccc}
\hline
GCVS name $(\log P)$& [Fe/H] & $M_\mathrm{V\,0}$ & $d$ & $R_0$       & $T_\mathrm{eff\,0}$ & $\mathfrak{M}$         & $\log g_\mathrm{stat}$ \\
\hskip5mm remark/ref.        &        & [mag]             & [pc]& [$R_\odot$] & [K]                 & [$\mathfrak{M}_\odot$] \\
\hline\hline
\multicolumn{4}{l}{RR Gem (-0.401)}\\
\hline
\hskip5mm A                             & \it{-0.15} & 0.68 & 1229 & 4.74 & 6882 & 0.80 & 2.99 \\
\hskip5mm B                             & \it{-0.15} & 0.66 & 1243 & 4.79 & 6881 & 0.82 & 2.99 \\
\hskip5mm C                             & \it{-0.15} & 0.67 & 1235 & 4.76 & 6884 & 0.81 & 2.99 \\
\hskip5mm D                             & \it{-0.15} & 0.65 & 1247 & 4.80 & 6884 & 0.83 & 2.99 \\
\hline
\hskip5mm \cite{lj90}                   & -0.20 & 0.99 &          1061 & 4.05 & 6699 & 0.55 & \it{2.96} \\
\hskip5mm \cite{fernley}                & -0.30 & 0.89 &      \it{1117} &      &      &      &      \\
\hskip5mm \cite{kovacs}                 & \it{-0.15} & 0.74 & \it{1196} & 4.60 & 6721 & \it{0.74} & \it{2.98} \\
\hskip5mm \cite{bono}                   & -0.29 & 0.92 &      \it{1100} \\
\hline\hline
\multicolumn{4}{l}{TT Lyn (-0.224)}\\
\hline
\hskip5mm A                             & \it{-1.20} & 0.86 & \ 626 & 5.32 & 6365 & 0.58 & 2.75 \\
\hskip5mm B                             & \it{-1.20} & 0.80 & \ 643 & 5.47 & 6365 & 0.62 & 2.75 \\
\hskip5mm C                             & \it{-1.20} & 0.75 & \ 660 & 5.62 & 6366 & 0.66 & 2.76 \\
\hskip5mm D                             & \it{-1.20} & 0.67 & \ 685 & 5.83 & 6366 & 0.73 & 2.77 \\
\hline
\hskip5mm \cite{lj90}                   & -1.35 & 0.75 &         \ 654 & 5.40 & 6284 & 0.62 & \it{2.77} \\
\hskip5mm \cite{fernley}                & -1.35 & 0.65 &\      \it{690} &      &      &      &      \\
\hskip5mm \cite{kovacs}                 & \it{-1.20} & 0.66 &\ \it{686} & 5.72 & 6283 & \it{0.69} & \it{2.76} \\
\hskip5mm \cite{bono}                   & -1.56 & 0.57 & \     \it{717} \\
\hline\hline
\multicolumn{4}{l}{RV Oct (-0.243)}\\
\hline
\hskip5mm A                             & \it{-1.15} & 0.32 & 1128 & 6.35 & 6567 & 0.97 & 2.82 \\
\hskip5mm B                             & \it{-1.15} & 0.28 & 1149 & 6.47 & 6566 & 1.02 & 2.82 \\
\hskip5mm C                             & \it{-1.15} & 0.43 & 1073 & 6.04 & 6565 & 0.85 & 2.81 \\
\hskip5mm D                             & \it{-1.15} & 0.40 & 1089 & 6.14 & 6565 & 0.89 & 2.81 \\
\hline
\hskip5mm A                             & -1.75 & 0.51 & 1038 & 5.94 & 6538 & 0.82 & 2.80 \\
\hskip5mm B                             & -1.75 & 0.43 & 1076 & 6.16 & 6537 & 0.90 & 2.81 \\
\hskip5mm C                             & -1.75 & 0.61 &\ 990 & 5.67 & 6537 & 0.73 & 2.79 \\
\hskip5mm D                             & -1.75 & 0.56 & 1014 & 5.81 & 6536 & 0.77 & 2.80 \\
\hline
\hskip5mm \cite{skillen}                & -1.75 & 0.76 &\         905 & 5.26 & 6437 & 0.62 & \it{2.79} \\
\hskip5mm \cite{fernley}                & -1.75 & 0.68 &   \ \it{ 956} &      &      &      &      \\
\hskip5mm \cite{bono}                   & -1.71 & 0.37 &     \it{1102} \\
\hline\hline
\multicolumn{4}{l}{AR Per (-0.371)}\\
\hline
\hskip5mm A                             & \it{-0.05} & 1.02 & \ 489 & 4.25 & 6719 & 0.54 & 2.92 \\
\hskip5mm B                             & \it{-0.05} & 1.03 & \ 487 & 4.23 & 6720 & 0.54 & 2.91 \\
\hskip5mm C                             & \it{-0.05} & 1.11 & \ 470 & 4.09 & 6715 & 0.49 & 2.91 \\
\hskip5mm D                             & \it{-0.05} & 1.11 & \ 469 & 4.08 & 6716 & 0.49 & 2.91 \\
\hline
\hskip5mm \cite{lj90}                   & -0.30 & 0.97 &          \ 500 & 4.15 & 6672 & 0.53 & \it{2.93} \\
\hskip5mm \cite{fernley}                & -0.30 & 0.87 &      \ \it{524} &      &      &      &      \\
\hskip5mm \cite{kovacs}                 & \it{-0.05} & 0.72 & \ \it{561} & 4.45 & 6833 & \it{0.61} & \it{2.93} \\
\hskip5mm \cite{bono}                   & -0.30 & 0.92 &  \     \it{512} \\
\hline\hline
\multicolumn{4}{l}{BB Pup (-0.318)}\\
\hline
\hskip5mm A                             & \it{-0.45} & 0.80 & 1648 & 4.65 & 6790 & 0.57 & 2.86 \\
\hskip5mm B                             & \it{-0.45} & 0.76 & 1677 & 4.73 & 6789 & 0.60 & 2.86 \\
\hskip5mm C                             & \it{-0.45} & 0.82 & 1633 & 4.61 & 6793 & 0.55 & 2.85 \\
\hskip5mm D                             & \it{-0.45} & 0.79 & 1658 & 4.67 & 6794 & 0.58 & 2.86 \\
\hline
\hskip5mm \cite{skillen}                & -0.60 & 1.21 &          1329 & 3.97 & 6551 & 0.40 & \it{2.84} \\
\hskip5mm \cite{fernley}                & -0.60 & 1.13 &      \it{1416} &      &      &      &      \\
\hskip5mm \cite{kovacs}                 & \it{-0.45} & 0.87 & \it{1596} & 4.40 & 6712 & \it{0.49} & \it{2.84} \\
\hskip5mm \cite{bono}                   & -0.64 & 0.65 &      \it{1766} \\
\hline\hline
\end{tabular}
\end{table*}

\begin{figure*} 
\begin{center}
\includegraphics[angle=0,width=5.8cm]{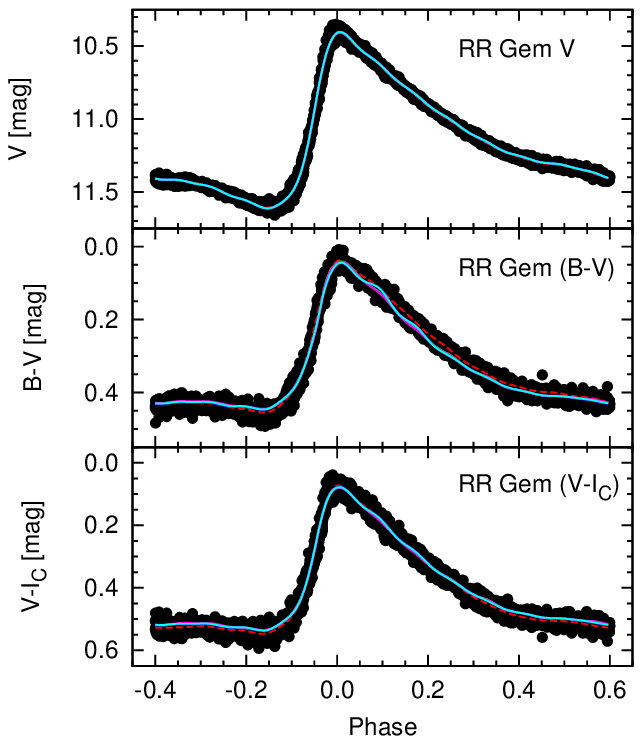}
\includegraphics[angle=0,width=5.8cm]{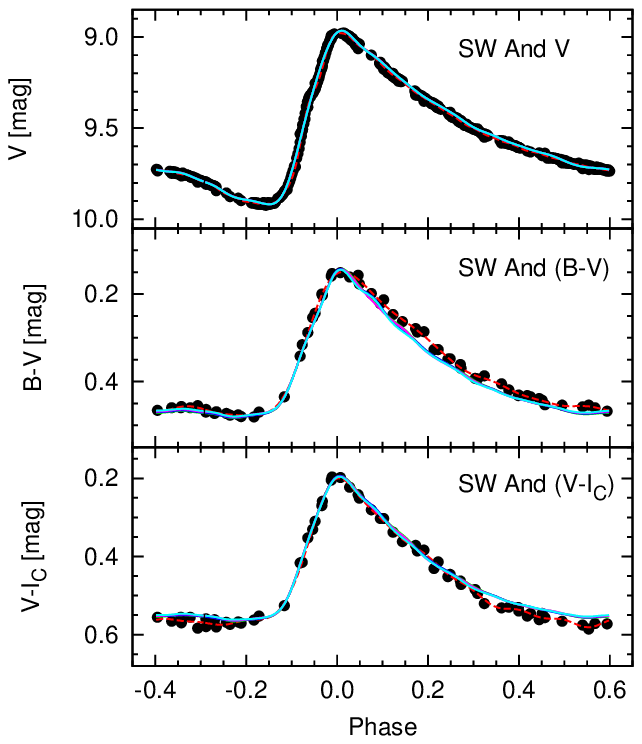}
\includegraphics[angle=0,width=5.8cm]{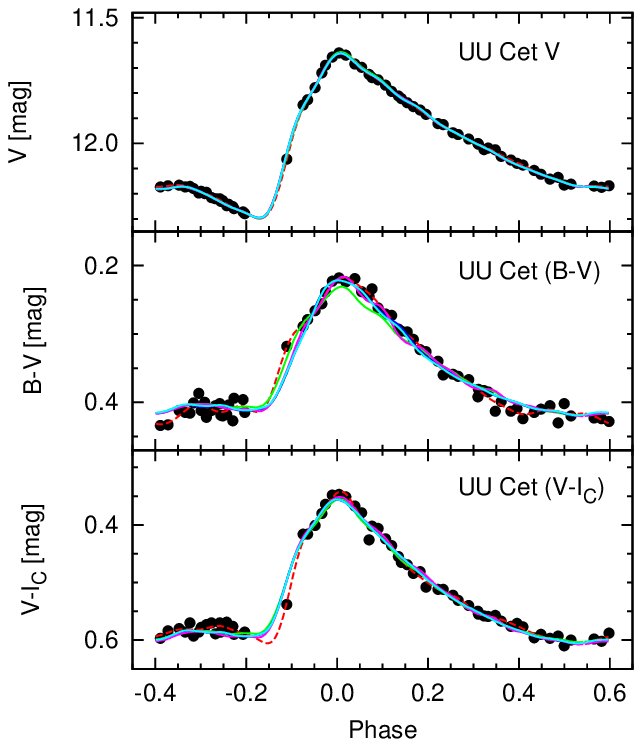}
\end{center}
\caption{Derived light curves (solid lines), input light curves (dashed line) and actual photometric measurements (dots) of RR~Gem, SW~And and UU~Cet, stars representing two better and a poorer quality input light curves, respectively. The four runs of the IP method with different settings are shown with different colours, however, the differences between these solutions are in most cases less than the thickness of the lines. \label{fig:lcfit}}
\end{figure*}
\begin{figure} 
\begin{center}
\includegraphics[angle=0,width=7cm]{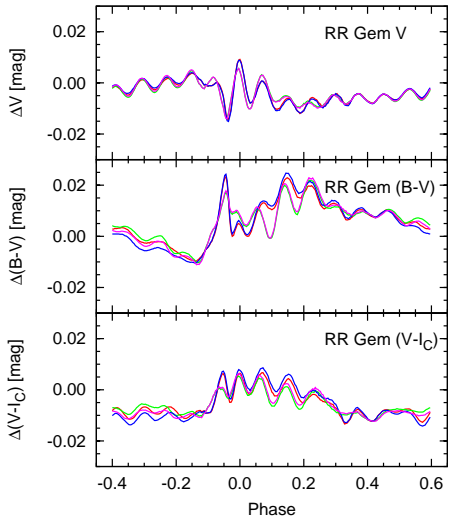}
\end{center}
\caption{Differences between the initial and solution light and colour curves for RR~Gem. Results of the IP method with four different settings are plotted with different colours. The periodic nature of the residuals arises from differences of the order of the Fourier sums used to describe the input and the fitted output curves.\label{fig:rrg-res}}
\end{figure}
\begin{figure*} 
\begin{center}
\includegraphics[angle=0,width=16.0cm]{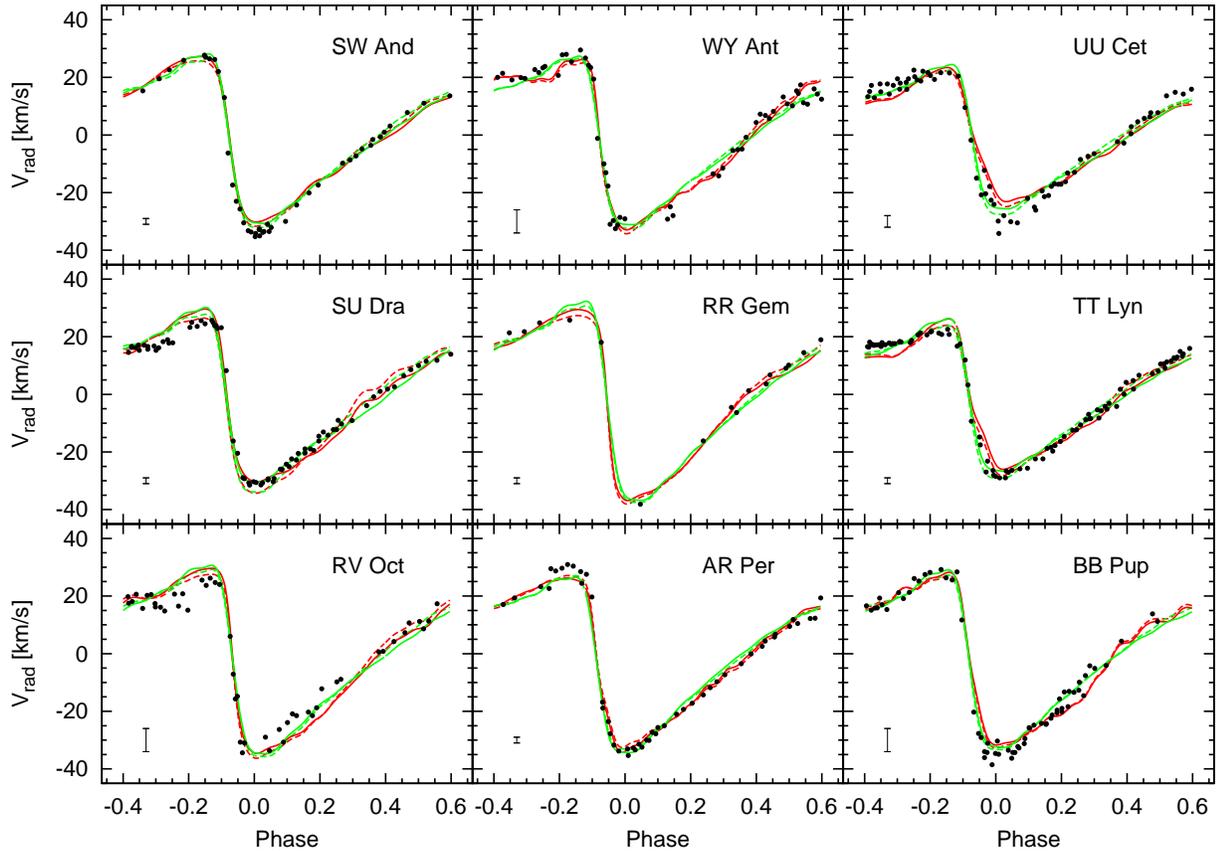}
\end{center}
\caption{Solutions for the $V_\mathrm{rad}(\varphi)$ curves of the 9 test objects are compared to their $V_\mathrm{rad}^{*}$ measurements. Gray (green) lines show Liu's template, black (red) lines mark the $V_\mathrm{rad}(\varphi)$ template according to Eq.~\ref{eq:i-vrad}. The medium weight solutions are drawn with solid lines, the large weight solutions are plotted with dashed lines. Dots show the actual $V_\mathrm{rad}^{*}$ measurements with mean values subtracted. Error bars in the bottom left corners of each panel indicate the mean errors of the observations. $V_\mathrm{rad}^{*}$ curves derived from $I_\mathrm{C}$ light curves are somewhat wavy when the light curve is of poorer quality and its high enough order Fourier fit is not smooth. Our $V_\mathrm{p}(\varphi)$ curves are transformed to $V_\mathrm{rad}(\varphi)$ using the actual value of the projection factor ($p$) given in the original papers. The centre-of-mass velocities of the observations are subtracted from the data. The slight differences between the observed radial velocity curves and the solution pulsation velocity curves transformed to radial velocities may either originate from the pulsation phase and period dependence of the transformation factor $p$ 
(see e.g. \citet{sabbey} and \citet{nardetto} for Cepheids) or from minor defects of the $V_\mathrm{rad}$ templates used.
\label{fig:9vrad}}
\end{figure*}
\begin{figure} 
\begin{center}
\includegraphics[angle=0,width=8cm]{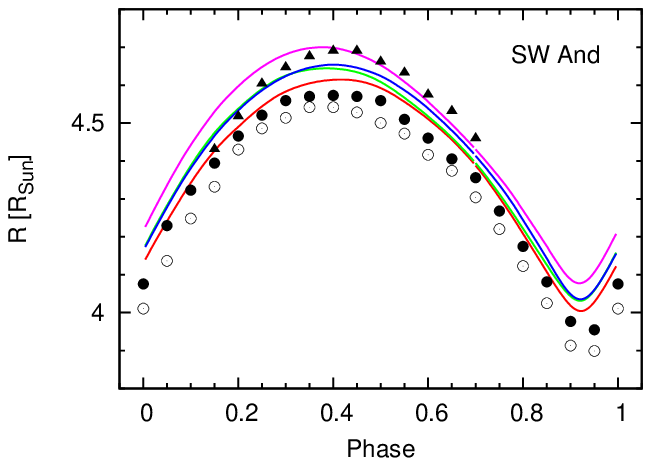}
\end{center}
\caption{Radius variation curves of SW~And derived from the IP method compared to results published by other authors. Continuous lines denote the solutions of the IP method with four different settings. Symbols denote literature data: triangles -- Cacciari et al. (1989),  filled circles -- Liu \& Janes (1990), open circles -- Jones et al. (1992).\label{fig:swand-r}}
\end{figure}

The IP method has been tested on 9 RRab stars. We have selected test objects that have good $BVI_\mathrm{C}$ light curves and $V_\mathrm{rad}$ observations available. All of these objects were previously subjected to \hbox{B-W} analyses by two or more authors. Four different runs of the IP method with different settings (detailed in Table~\ref{tbl:abcd}) have been performed for each test object to make uncertainity estimations possible. These settings are marked with the letters A, B, C, and D in the `remark/ref.' columns of Tables~\ref{tbl:res} and \ref{tbl:mod-test}. Two different $V_\mathrm{p}(\varphi)$ template curves (see Sects. \ref{sect:liuvrad}, \ref{sect:i-vrad}) have been applied with medium and large $V_\mathrm{p}(\varphi)$ weight factors.

The results of the tests are compared to the results of earlier \hbox{B-W} analyses and to the $M_\mathrm{V\,0}$ absolute brightness values derived from pulsational luminosities \citep{bono} in Table~\ref{tbl:res}. This table lists the source of the data, the metallicity, mean absolute visual magnitude ($M_\mathrm{V\,0}$), distance ($d$), mean radius ($R_0$), mean effective surface temperature ($T_\mathrm{eff\,0}$), mass ($\mathfrak{M}$), and static gravity ($\log g_\mathrm{stat}$) of the stars. If the mass, $\log g_\mathrm{stat}$, and/or distance were not given in the original paper cited, they were calculated according to Eqs.~\ref{eq:puls}, \ref{eq:logg}, and \ref{eq:logd}, respectively. These values are typeset with italics in Table~\ref{tbl:res}. The static gravity and distance are expressed as:

\begin{equation}
\log g_\mathrm{stat} = \log \mathfrak{M} - 2 \log R_\mathrm{0} + 4.438. \label{eq:logg}
\end{equation}

\noindent and

\begin{equation}
\log d = 0.2 \cdot (m_\mathrm{V\,0} - M_\mathrm{V\,0} - A_\mathrm{V} + 5) \label{eq:logd}
\end{equation}

\noindent where $m_\mathrm{V\,0}$ is the mean apparent $V$ magnitude and $A_\mathrm{V}$ is the interstellar extinction in $V$ band.

The original \hbox{B-W} analyses used spectroscopic metallicity values except \cite{kovacs} who calculated the [Fe/H] of the studied variables from the light curve Fourier parameters \citep[][Eq.~3, photometric metallicity]{jk96}. In Table~\ref{tbl:res} [Fe/H] values corresponding to this formula are typeset with italics. As the photometric metallicities were calibrated on spectroscopic data the two metallicities are identical within the limits of the uncertainties. For all stars, with the exception of RV~Oct, the spectroscopic and photometric metallicities agree within $0.0 - 0.3$\,dex. The spectroscopic metallicity of RV~Oct is $-1.75$\,dex, while photometrically $-1.15$\,dex is derived. Therefore, we performed all the 4 runs of the method with both metallicity values for this star. For other test objects we accepted the photometric metallicities. The metallicities used by \cite{bono} are taken from the compilation by \cite{fernley98}. We have also checked how the results change if $\alpha$ enhanced models are used. The IP method was run for SU~Dra, for one of the metal poor stars using ${\rm [M/H]=-1.6}$, ${\rm [\alpha/Fe]=0.4}$ atmosphere models, too. In Table~\ref{tbl:res} these solutions are denoted by `$\alpha$'. The effect of using $\alpha$ enhanced models are small, it is about $+5$\,K and $-0.04$\,mag in $T_\mathrm{eff}$ and $M_\mathrm{V}$, respectively. For compatibility reasons (all the previous direct B-W analyses used solar scaled atmosphere models) in the following, solutions of the IP method using solar scaled atmosphere models are regarded.

The derived light curves reproduce the input light curves to an accuracy of $0.01-0.03$\,mag which is about the observational uncertainty as shown in Figs~\ref{fig:lcfit} and \ref{fig:rrg-res}. The results of the fitting process are shown by the examples of SW~And and UU~Cet in Fig.~\ref{fig:lcfit}. The input light curves of SW~And are well defined by the observations, especially in $V$ band, while UU~Cet has poorer photometric data, particularly at the lower part of the rising branch. The method produces more stable result (i.e., the difference between the four solutions are smaller) for those stars that have better quality light curves. Fig.~\ref{fig:rrg-res} shows the difference between the input and derived light and colour curves of RR~Gem. The differences are less than about $0.01$\,mag in $V$ and $(V-I_\mathrm{C})$ and less than $0.02$\,mag in $(B-V)$.

As a result of the IP method, solution $V_\mathrm{p}(\varphi)$ curves are also derived, which can be converted to $V_\mathrm{rad}(\varphi)$ curves. These curves are compared to the actual $V_\mathrm{rad}$ measurements of our 9 test objects in Fig.~\ref{fig:9vrad}. Our $V_\mathrm{p}(\varphi)$ curves were transformed to $V_\mathrm{rad}(\varphi)$ using the $p$ transformation factors given by authors of the measurements. The radial velocity data have typically 1-4 km/s uncertainties, as error bars indicate in the plots. Data are taken from \citet[SW~And]{cacc87}, \citet[UU~Cet]{clementini}, \citet[SU~Dra, RR~Gem, TT~Lyn, AR~Per]{lj89} and \citet[WY~Ant, RV~Oct, BB~Pup]{skillen}. Note that the $V_\mathrm{rad}$ observations were not taken into account by the IP method in any way. In most cases the radial velocity curves derived from the IP method agree within the uncertainties with the observations. There are some systematic differences within about 5\,km/s range apparent, most probaly due to minor defects of both Liu's $V_\mathrm{rad}$ template and the $V_\mathrm p-I_\mathrm C$ ralation. The largest deviations are observed at around maximum brightness phases of UU~Cet. However, the good agreement between the physical parameters derived even for this star from the IP method and from direct B-W analyses shows that our method is not sensitive to these small inaccuracies of the solutions for the radial velocity curves.

From the solution $V_\mathrm{p}(\varphi)$ curve and from the mean radius the variation of the radius with phase, $R(\varphi)$, can be derived. The four solution $R(\varphi)$ curves of SW~And are compared to the curves given by authors of earlier \hbox{B-W} analyses in Fig.~\ref{fig:swand-r}. The agreement here is fully satisfactory.

\begin{figure} 
\begin{center}
\includegraphics[angle=0,width=8cm]{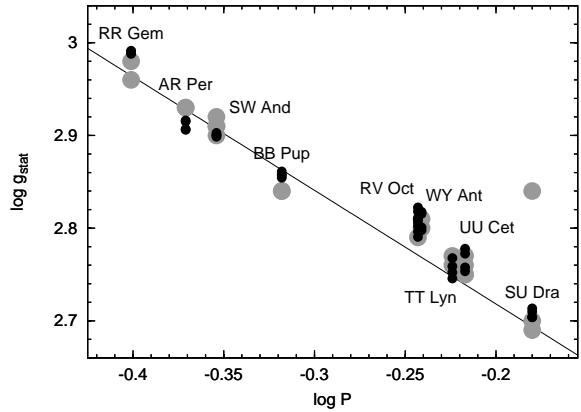}
\end{center}
\caption{$\log P$ vs.\ $\log g_\mathrm{stat}$ values from Table~\ref{tbl:res}. The results of the IP method are plotted with black dots while literature data are marked with gray colour. The scatter of the literature data and the data from the four runs of the IP method with different settings indicate the inherent uncertainties of direct B-W analyses and of the IP method. Solid line corresponds to the empirical formula given by \citet[Eq.~15]{jurcsik98}.\label{fig:logp-logg}}
\end{figure}

\begin{figure} 
\begin{center}
\includegraphics[angle=0,width=8cm]{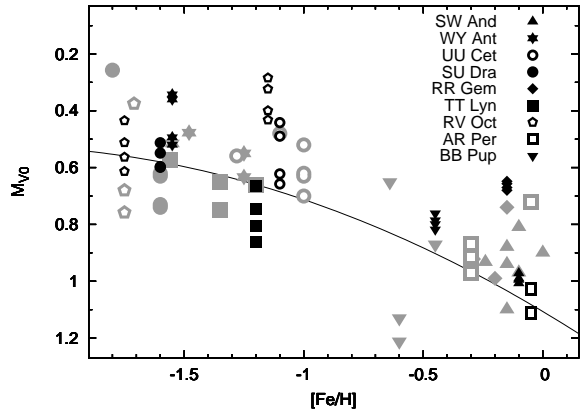}
\end{center}
\caption{[Fe/H] vs.\ $M_\mathrm{V\,0}$ values from Table~\ref{tbl:res}. Our results are marked with black symbols while literature data are plotted with gray colour using the same symbols. Solid line shows the zero age horizontal branch relation given by \citet[Eq.~9]{st}.\label{fig:fe-mv}}
\end{figure}

\begin{figure} 
\begin{center}
\includegraphics[angle=0,width=8cm]{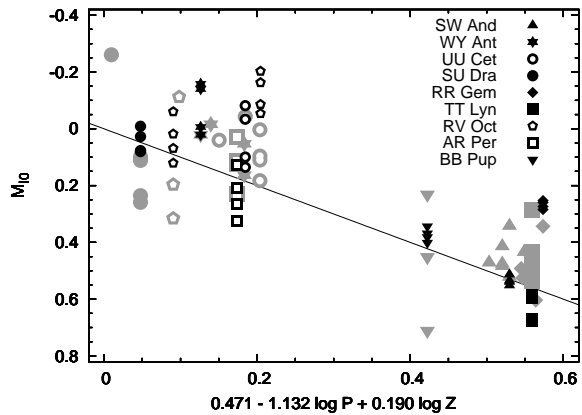}
\end{center}
\caption{Absolute $I_{\rm C}$ brightness values ($M\mathrm{I\,0}$) derived from the IP method (black symbols) and from direct B-W analyses (grey symbols) compared to their predicted values from the $M_{\mathrm I\,0}(\log P, \log Z)$ relation defined by \citet[Eq.~3]{catelan}. Solid line shows the $x=y$ equality.
\label{fig:Mipz}}
\end{figure}

\begin{figure} 
\begin{center}
\includegraphics[angle=0,width=8cm]{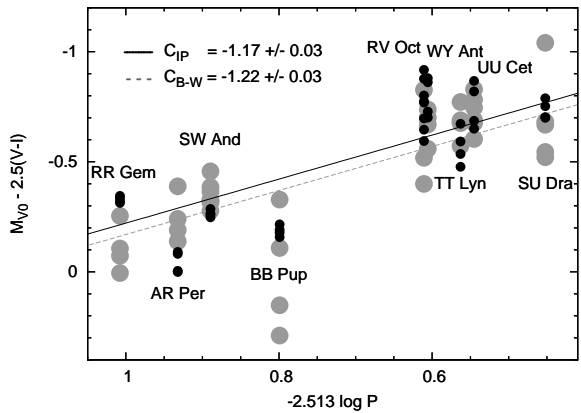}
\end{center}
\caption{ Period--$V$ brightness--$(V-I_{\rm C})$ colour realtion according to Eq.~8 of \citet{kw}. The constant of this equation ($C$) was calculated by least squares fits to our data (black dots) and for the literature data (grey dots). Within the uncertainties the same constant of the PLC($V$, $I_{\rm C}$) relation was obtained from both data sets with the same $r.m.s.$ scatter of $0.03$\,mag. Solid and dashed lines correspond to the $y=x+C$ equalities for IP and literature B-W data, respectively. \label{fig:plc}}
\end{figure}

The physical parameters derived from the IP method are in reasonably good agreement with results of other methods, as shown in Table~\ref{tbl:res}. For individual stars the $M_\mathrm{V\,0}$ values derived by different authors differ by about $0.1-0.3$\,mag, while the different settings of the IP method give results within $0.2$\,mag range. As there is no objective reason to decide which of the literature data are the most correct, the direct comparison of our results with them is not obvious. E.g., the absolute magnitudes derived from the IP method for BB~Pup agree well with the value given by \cite{kovacs}, while \cite{skillen} and \cite{fernley} give $0.2-0.3$\,mag fainter results.

Therefore, we decided to check the consistency of our results with empirical and theoretical relations valid for RR~Lyrae variables in order to become certain about the reliability of our results.

There is a well-defined relation between the $\log P$ and $\log g$ values of pulsating variables \citep[see e.g.][]{fernie}. The $\log P$ vs.\ $\log g_\mathrm{stat}$ points of the IP method solutions and the data given by other authors for the test objects are plotted in Fig.~\ref{fig:logp-logg}. Solid line shows the empirical relation given by \citet[Eq.~15]{jurcsik98}. Our results fit to the empirical relation similarly well as the data of direct \hbox{B-W} analyses.

Another important relation exists between the metallicity and the average absolute visual brightness of Horizontal Branch stars. The [Fe/H] vs.\ $M_\mathrm{V\,0}$ results of the IP method and of other authors are compared to zero age horizontal branch (ZAHB) model predictions \citep{st} in Fig.~\ref{fig:fe-mv}. This figure also shows that our results are in good accordance with theoretical predictions. The ambiguity of our results arising from the different settings of the IP method is smaller for most of the stars than the ambiguity between the results of earlier analyses.

The reliability of the results was also checked and compared with direct B-W data using PL and PLC relations of the absolute $I_\mathrm C$ brightness as shown in Figs. \ref{fig:Mipz} and \ref{fig:plc}. The $M_\mathrm I (\log P, \log Z)$ and the $\log P(I_\mathrm C,V-I_\mathrm C)$ relations defined by \cite{catelan} and \cite{kw} are used for these tests. Again, no discrepancy of the results of the IP method are evident.

\section{Applicability}

We have also tested how the IP method works if not all the necessary input data are available. The results are shown in Table~\ref{tbl:mod-test} for SW~And and RR~Gem, the two stars that have the best quality photometric observations among the test objects.

\begin{table*} 
\caption{Results of the applicability tests on SW~And and RR~Gem. The two-colour ($VI_\mathrm{C}$ and $BV$) solutions are marked with `vi' and `bv', respectively, and solutions when the zero points of the colours are also fitted are marked with `zp' in the `remark' column.
\label{tbl:mod-test}}
\begin{tabular}{lp{15mm}crcccccc}
\hline
GCVS name $(\log P)$& [Fe/H] & $M_\mathrm{V\,0}$ & $d$ & $R_0$       & $T_\mathrm{eff\,0}$ & $\mathfrak{M}$         & $\log g_\mathrm{stat}$ & $\Delta(B-V)_0$ & $\Delta(V-I)_0$ \\
\hskip5mm remark$^*$              &        & [mag]             & [pc]& [$R_\odot$] & [K]                 & [$\mathfrak{M}_\odot$] & & [mag] & [mag] \\
\hline\hline
\multicolumn{4}{l}{SW And (-0.354)}\\
\hline
\hskip5mm A vi          & \it{-0.10} & 0.92 &\ 527 & 4.63 & 6609 & 0.64 & 2.91 \\
\hskip5mm B vi          & \it{-0.10} & 0.91 &\ 529 & 4.65 & 6609 & 0.65 & 2.91 \\
\hskip5mm C vi          & \it{-0.10} & 0.89 &\ 534 & 4.69 & 6609 & 0.66 & 2.91 \\
\hskip5mm D vi          & \it{-0.10} & 0.91 &\ 531 & 4.66 & 6608 & 0.65 & 2.91 \\
\hline
\hskip5mm C bv          & \it{-0.10} & 1.14 &\ 475 & 4.08 & 6681 & 0.46 & 2.88 \\
\hskip5mm D bv          & \it{-0.10} & 1.13 &\ 479 & 4.11 & 6680 & 0.47 & 2.88 \\
\hline
\hskip5mm A bvi zp      & \it{-0.10} & 0.95 &\ 520 & 4.42 & 6707 & 0.57 & 2.90 &\ 0.005 & 0.027 \\
\hskip5mm B bvi zp      & \it{-0.10} & 0.98 &\ 512 & 4.41 & 6665 & 0.56 & 2.90 & -0.005 & 0.016 \\
\hskip5mm C bvi zp      & \it{-0.10} & 0.93 &\ 525 & 4.47 & 6705 & 0.58 & 2.90 &\ 0.005 & 0.026 \\
\hskip5mm D bvi zp      & \it{-0.10} & 0.92 &\ 527 & 4.48 & 6708 & 0.59 & 2.90 &\ 0.006 & 0.027 \\
\hline
\hskip5mm A vi zp       & \it{-0.10} & 0.69 &\ 584 & 4.75 & 6842 & 0.68 & 2.92 &        & 0.062 \\
\hskip5mm B vi zp       & \it{-0.10} & 0.70 &\ 583 & 4.75 & 6841 & 0.68 & 2.92 &        & 0.061 \\
\hskip5mm C vi zp       & \it{-0.10} & 0.68 &\ 589 & 4.80 & 6840 & 0.70 & 2.92 &        & 0.061 \\
\hskip5mm D vi zp       & \it{-0.10} & 0.67 &\ 591 & 4.81 & 6842 & 0.70 & 2.92 &        & 0.062 \\
\hline\hline
\multicolumn{4}{l}{RR Gem (-0.401)}\\
\hline
\hskip5mm A vi          & \it{-0.15} & 0.80 & 1168 & 4.52 & 6858 & 0.71 & 2.98 \\
\hskip5mm B vi          & \it{-0.15} & 0.79 & 1171 & 4.54 & 6856 & 0.71 & 2.98 \\
\hskip5mm C vi          & \it{-0.15} & 0.77 & 1183 & 4.58 & 6859 & 0.73 & 2.98 \\
\hskip5mm D vi          & \it{-0.15} & 0.77 & 1183 & 4.58 & 6859 & 0.73 & 2.98 \\
\hline
\hskip5mm C bv          & \it{-0.15} & 0.40 & 1400 & 5.34 & 6914 & 1.08 & 3.02 \\
\hskip5mm D bv          & \it{-0.15} & 0.40 & 1401 & 5.34 & 6914 & 1.08 & 3.02 \\
\hline
\hskip5mm A bvi zp      & \it{-0.15} & 0.66 & 1241 & 4.74 & 6907 & 0.80 & 2.99 &\ 0.000 & 0.012 \\
\hskip5mm B bvi zp      & \it{-0.15} & 0.63 & 1258 & 4.83 & 6890 & 0.84 & 2.99 & -0.004 & 0.008 \\
\hskip5mm C bvi zp      & \it{-0.15} & 0.63 & 1259 & 4.76 & 6940 & 0.81 & 2.99 &\ 0.007 & 0.019 \\
\hskip5mm D bvi zp      & \it{-0.15} & 0.64 & 1255 & 4.81 & 6893 & 0.83 & 2.99 & -0.004 & 0.008 \\
\hline
\hskip5mm A vi zp       & \it{-0.15} & 0.74 & 1199 & 4.52 & 6946 & 0.71 & 2.98 &        & 0.022 \\
\hskip5mm B vi zp       & \it{-0.15} & 0.72 & 1205 & 4.51 & 6973 & 0.70 & 2.98 &        & 0.029 \\
\hskip5mm C vi zp       & \it{-0.15} & 0.73 & 1202 & 4.57 & 6918 & 0.73 & 2.98 &        & 0.015 \\
\hskip5mm D vi zp       & \it{-0.15} & 0.77 & 1183 & 4.57 & 6865 & 0.73 & 2.98 &        & 0.002 \\
\hline\hline
\multicolumn{10}{p{135mm}}{
$^*$: On the notation A, B, C, and D see Table~\ref{tbl:abcd}}
\end{tabular}
\end{table*}

\subsection{Two colour ($VI_\mathrm{C}$ or $BV$) light curves}

The IP method was tested using only $V$ and $I_\mathrm{C}$ or $B$ and $V$ light curves. As the initial $T_\mathrm{eff}(\varphi)$ and $V_\mathrm{p}(\varphi)$ curves were defined in Sect. 2 partially using the $I_\mathrm{C}$ data, if only $BV$ data are available the initial $T_\mathrm{eff}(\varphi)$ curve is defined from empirical $T_\mathrm{eff}(B-V,[\mathrm{Fe/H}])$ relation \cite[Table 2., Eq. 3 in][]{alonso}, and in this case, Liu's template  $V_\mathrm{p}(\varphi)$  curve can only be used.

The results of four and two runs with different settings taking only $V$ and $I_\mathrm{C}$ or $B$ and $V$ light curves of SW~And and RR~Gem into account are given in Table~\ref{tbl:mod-test} marked with `vi' and `bv' in the `remark' column.

The IP method using only $V$ and $I_\mathrm{C}$ light curves gives $0.1$\,mag brighter and fainter $M_\mathrm{V\,0}$ values, $0.2\,R_\odot$ larger and smaller mean radii, and $0.1\,\mathfrak{M}_\odot$ greater and smaller masses than the $BVI_\mathrm{C}$ solutions for SW~And and RR~Gem, respectively, while the temperature of both stars are cooler by $20-40$ K according to these solutions. These values are not significantly discrepant from literature data, the agreement for RR~Gem is even better than when using $BVI_C$ data.

If only $B$ and $V$ data are utilized the temperature solution of both stars are hotter by $30-40$\,K. In this case the other parameters of SW And, though are still resonable, differ from the direct B-W results more than the results of the $BVI_\mathrm C$ solutions. The $BV$ solutions give, however, unreliably bright $M_{\mathrm V\,0}$ and large radius and mass for RR Gem.

Therfore, we conclude that the IP method give acceptable solutions only if at least $V$ and $I_\mathrm C$ band light curves are available, but for $B$ and $V$ data it may lead to unreliable results. In direct \hbox{B-W} analyses it was also found by many authors that light curves observed in longer wavelength passbands ($VI_\mathrm CJHK$) give more accurate results than those observed in bluer passbands ($UB$).

\subsection{Unknown zero points of the colour curves}\label{sect:app-zp}

To fix the zero points of the light and colour curves the reddening and standard magnitudes of the variable and/or the comparison stars have to be known. However, these conditions are not always met.

We have also performed test runs to check the results of the IP method when the $(B-V)_0$ and $(V-I_\mathrm{C})_0$ zero points are not known. During these tests the colour curves' zero points were not fixed but they were also fitted by the algorithm. Our tests show that, for good quality light curves, the algorithm converges to the appropriate zero points. Even when the initial values of the zero points of the colour curves are off about 0.1\,mag from their correct values, the method finds them within about $0.01-0.03$\,mag accuracy. These differences cause only marginal changes in $M_\mathrm{V\,0}$, $R$, and $,\mathfrak{M}$, 0.06\,mag, 0.06\,$R_\odot$, and 0.02\,$\mathfrak{M}_\odot$, respectively. The output of these four runs are listed in Table~\ref{tbl:mod-test}, marked with `zp' in the `remark' column. For these results the zero point differences are also given in the last two columns with respect to the observed zero points ($\Delta (B-V)_0$ and $\Delta (V-I_\mathrm{C})_0$).

\vskip12pt

Finally, we have also checked what results are obtained if neither $B$ observations, nor $(V-I_\mathrm{C})_0$ zero point were available. These results are denoted by `vi~zp' in Table~\ref{tbl:mod-test}. For the two stars even these solutions remain reliable. We have to emphasize, however, that in the case of incomplete input data the quality of the observed light curves becomes even more important. For poorer quality light curves these limited solutions may lead to physically meaningless results.

\section{Summary and conclusions}

We introduced a new Inverse Photometric Baade-Wesselink method to determine physical parameters of RRab stars exclusively from their multicolour light curves. It was shown through the example of 9 test objects that for good quality $BVI_\mathrm{C}$ light curves the IP method gives similarly accurate results to other analyses, without the need for spectroscopic $V_\mathrm{rad}$ observations.

The success of the IP method depends strongly on the quality of the light curves. For the best quality input curves the method is applicable even when only $V$ and $I_\mathrm{C}$ observations are available and when the zero points of the colours are not known. Note that any error in $m_\mathrm{V\,0}$ influences only the distance determination.

The method is an inverse method, since the primary curves are the $V_\mathrm{p}(\varphi)$ and $T_\mathrm{eff}(\varphi)$ functions and the output light curves are derived from them through the application of static atmosphere models, while direct \hbox{B-W} methods derive $T_\mathrm{eff}(\varphi)$ and $\log g_\mathrm{eff}(\varphi)$ curves directly from the observations. The IP method varies these primary curves during a nonlinear least squares fitting process so that the derived light and colour curves best match the observed input curves. The initial $T_\mathrm{eff}(\varphi)$ curve is calculated from the $(V-I_\mathrm{C})(\varphi)$ colour curve using an empirical formula \citep[][Table 8.]{bessel}. The initial $V_\mathrm{p}(\varphi)$ curve is defined twofold. We have derived a new \hbox{$I_\mathrm{C}$ -- $V_\mathrm{p}$} relation which yields the $V_\mathrm{p}(\varphi)$ curve from the $I_\mathrm{C}(\varphi)$ light curve with an accuracy of 3.5\,km/s (Eq.~\ref{eq:i-vrad}). Alternatively, the $V_\mathrm{p}(\varphi)$ template of \cite{liu91} is used.

The method was developed to extract physical information from accurate multicolour RR~Lyrae light curves. During the development of the IP method special attention was payed to its applicability to RR~Lyrae stars showing Blazhko modulation. The applicability and the results of the method for a strongly modulated RR~Lyrae star, MW~Lyr, is shown in \cite{mwlyr2}. The IP method is the only possibility today to derive the variations of the physical parameters of a Blazhko star throughout the modulation cycle.

Another field of interest of the application of this new method is to study RRab stars in globular clusters or dwarf galaxies since there are very few radial velocity observations made of these faint variables but, thanks to the CCD technique, accurate multicolour data are available.

The IP method has been specialised to fundamental mode RR~Lyrae stars, although with the appropriate choice of the initial $V_\mathrm{p}(\varphi)$ curve and using adequate pulsation equation it could be adapted for other type of radial pulsators. Using appropriate model atmosphere tables, the method is applicable for other passbands, as well. To do this, however, empirical formulae between the appropriate colours and brightnesses, and the $V_\mathrm{p}(\varphi)$ and $T_\mathrm{eff}(\varphi)$ curves have to be defined first.

\section*{Acknowledgments}

We thank to an anonymous referee for his/her very helpful comments and suggestions from which our paper benefited a lot. We are grateful to G\'eza Kov\'acs for helpful discussions and for the photometric and $V_\mathrm{rad}$ data he kindly provided us in electronic form. We thank J\'anos Nuspl for the useful discussions on atmosphere models and Istv\'an D\'ek\'any for his help at the beginning of this work. The financial support of OTKA grants T-068626 and T-048961 is acknowledged.

\bsp
\label{lastpage}

\end{document}